\documentclass[twocolumn,showpacs,showkeys,aps,pra,floatfix]{revtex4-1}
\pdfoutput=1
\usepackage{amsmath}
\usepackage{graphicx}
\usepackage{dcolumn}
\begin{document}

\title{Exploring the foundations of quantum mechanics using 
Monte Carlo simulations of the Freedman-Clauser experimental
test of Bell's Inequality}

\author{S.D. Foulkes}
\email[]{sfoulkes2@gmail.com}

\date{5 April 2013}

\begin{abstract}
Monte Carlo simulations of the Freedman-Clauser experiment
are used to test the generic wave function collapse 
model of Quantum Mechanics, a local realistic model,
and a dynamical state reduction model of wave function collapse.
The simulated results are compared to the 
actual results of the experiment which confirmed 
the quantum mechanical calculation
for nine different relative angles between the two
polarization analyzers.  
For each simulation $5\times10^7$ total simulated photon pairs 
were generated at each relative angle. The generic 
wave function collapse model closely followed the general 
shape of the theoretical calculation but differed from
the calculated values by $2.5\%$ to $3.3\%$ for angles less than
or equal to $\pi/8$ and differed by $15.0\%$ to $52.5\%$ for
angles greater than or equal to $3\pi/8$.  
The local realistic model did not
replicate the experimental results but was generally
within $1\%$ of a classical calculation for all
analyzer angles. A dynamical state reduction/collapse model, 
approximated by using a ``smeared'' polarization, 
yielded values within $1\%$ of
the quantum mechanical calculation and provides
an independent estimate of the correlation length
used in these models of  $r_c = (1.04 \pm 0.14)\times 10^{-5}$ cm.

\end{abstract}


\maketitle
%My document startes here
% Background section, wave function collapse Inter., Bohr, and EPR
\section{Introduction}
\label{sec-intro}
Quantum mechanics (QM) is one of the most successful theories in the 
history of science.  Since it's development in the early
twentieth century, the predictions of QM have
been repeatedly confirmed by experiment to extremely high
precision.  The physical interpretation of what the
mathematical formalism represents, however, has been the
subject of much debate.  
Most physicists simply apply the formalism and are either
not concerned with a physical interpretation
or accept the Copenhagen interpretation
as reasonable.  Exactly what constitutes the
Copenhagen interpretation, however, is not always
completely clear.  According to Faye~\cite{faye:2008}
the Copenhagen interpretation 
is generally related to indeterminism, a statistical
interpretation of the wave function, and
Bohr's concept of complementarity.
To most physicists, however, it also includes the collapse
of the wave function during a measurement.  This leads
to the ``measurement problem'' which is generally defined
as the conflict between the linear dynamics of
QM and the non-linear collapse of
the wave function during a measurement. 

In their 1935 paper Einstein, Podolsky, and Rosen (EPR)
~\cite{EPR:1966}
argued that QM was incomplete.
In their analysis EPR distinguish between 
``physical reality''
and the ``physical concepts'' of a theory that are 
intended to correspond to physical reality.
EPR did not attempt to extensively 
define ``physical reality''
but rather asserted that - ``If, without in any way disturbing a system,
we can predict with certainty (i.e., with probability equal to unity)
the value of a physical quantity, then there exists an element of physical
reality corresponding to this physical quantity''.  

While not specifically stated, 
EPR tacitly assume that ``elements of physical
reality'' have precise values. The physical quantities predicted by a
particular theory and measured by a particular experiment contain
a level of uncertainty, but the underlying physical reality is assumed 
to be an exact quantity independent of any theory or measurement.  
This assumption is critical to their analysis for without it, 
no theory would be able to meet the above criterion which they 
believe is both a ``reasonable'' and ``sufficient'' way of 
``recognizing a physical reality''.  

EPR also require locality and admit that their conclusion would
not be valid if non-locality is allowed.  However, they reject
non-locality stating ``no reasonable definition of reality
could be expected to permit this''.
Since QM precludes precise simultaneous determination 
of two physical quantities with 
non-commuting operators(e.g., position and momentum), 
EPR conclude QM must be incomplete.

Later authors attempted to resolve the EPR paradox by postulating
so called ``hidden variables'' theories. 
J.S. Bell
~\cite{bell:1964}
made the issue more explicit and
proved an inequality for what he termed ``local realistic 
theories'' that is violated
by QM calculations for an entangled state of a composite system.
Bell's work provided an experimental framework to resolve
the apparent conflict between local realistic or hidden variables
theories and QM.  

A specific experiment to test Bell's inequality was proposed by 
Clauser, Horne, Shimony, and Holt
~\cite{chsh:1969} in 1969.  
The proposed experiment measured the correlation in linear
polarization of two entangled
photons emitted in a $J=0 \rightarrow J=1 \rightarrow J=0$
atomic cascade of Calcium.  

The proposed experiment was carried out by Freedman and Clauser in 1972
~\cite{freedman-clauser:1972}
for nine different relative angles between their polarization analyzers and 
confirmed, within the experimental error, the QM calculation at each angle.
The results of Freedman and Clauser have since been confirmed
by several other experiments
~\cite{aspect:1982,weihs:1998}, (for complete review
of experiments related to the Bell inequality see ~\cite{shimony:2009}).

The results of Bell inequality experiments have lead to a general
acceptance of the non-local nature of QM, but have not resolved
the measurement problem or the mechanism of wave function collapse.
One avenue of research addressing this issue is Collapse Theory
(dynamical reduction models) which
attempts to modify the dynamical equations of QM to address the transition
from microscopic to macroscopic objects~\cite{ghirardi:1986,grassi:1990}
(for a review of dynamical reduction models see~\cite{bassi:2003} and 
the most recent progress on a relativistic state reduction model 
see~\cite{bedingham:2011}). As yet no definitive experimental test of these
models has been performed. % ******************  check *****************

It is important to note that the QM calculation 
tested by the Freedman-Clauser experiment and others does not
require physical reality to have precise values nor does it require
non-locality.  The QM calculation simply begins with an assumed
initial state of the system and applies the mathematical 
techniques set forth in QM to determine the probabilities associated
with the possible final states
that are expected to occur in the experiment. 
It is the physical interpretation
of QM that leads to non-locality and a generally accepted concept
of physical reality that leads to precise values of physical quantities.

The focus of this analysis is to evaluate, using Monte Carlo simulations
of the Freedman-Clauser experiment, the generic non-local collapse of the 
wave function interpretation of QM, 
a local realistic interpretation,
and an alternative physical interpretation wherein
physical reality is not assumed to include precise values
for physical quantities.  The latter interpretation
is intended to simulate dynamical state reduction/collapse models where 
physical quantities are ``smeared'' using Bedingham's 
terminology~\cite{bedingham:2011}.

Dynamical state reduction/collapse models add
a stochastic term into the dynamical equation of QM 
corresponding to the reduction process (for a review
of wave-function collapse models see~\cite{bassietal:2012}).
This process is ``formally identical to an approximate position 
measurement''~\cite{ghirardi:1986}.  Early work on
dynamical state reduction/collapse models introduced
the concept in this way ~\cite{barchielli:1982,barchielli:1983}.
In his relativistic 
state reduction model Bedingham~\cite{bedingham:2011}
utilizes a mediating field (called a ``pointer field'') to
``smear'' the interactions. 

To approximate dynamical state reduction/collapse models
we therefore evaluate a model where the photon polarization 
is not assumed to have a precise value but rather
is random with a Gaussian distribution about an appropriate mean. 

We first utilize the generic wave function collapse 
interpretation of QM, 
with it's inherently non-local characteristics, as the
physical concept and demonstrate it provides a close,
but not precise, match to the QM calculation.

We then show that a
``local realistic'' physical concept is completely inconsistent with
the QM calculation and the results of experiment, 
but rather is in close agreement with a classical 
calculation of the expected coincidence rate. 

We then remove the assumption that physical reality
contains precise values for physical quantities
and evaluate the dynamical state reduction/collapse model approach
and find that it more accurately corresponds to the predictions
of QM that were confirmed by the Freedman-Clauser
experiment and others.

% calculation of the probability of a gamma
% passing an analyzer with given efficiencies
% QM calculation

\section{QM calculation of the coincidence rate}
\label{sec-qm_calc}
Following the procedure in Horne ~\cite{horne:1970}
the probability of transmitting a linearly polarized
photon using a real analyzer with known efficiencies
is given by the following efficiency matrix
$$
P = \left( \begin{array}{cc}
     \epsilon_{\parallel}  &  0                \\
           0               &  \epsilon_{\perp}
     \end{array} \right)
$$
where $\epsilon_{\parallel}$ is the probability of
transmitting a photon with linear polarization
parallel to the analyzer and $\epsilon_{\perp}$ is
the probability of transmitting a photon polarized
perpendicular to the analyzer (leakage). 
Transforming to a
basis for a photon with an arbitrary angle of 
polarization, $\phi$, the probability
of transmission for both QM and classical 
calculations becomes
% equation for the probability of transmission
\begin{equation}
P(\phi)=\epsilon_{\parallel}\cos^{2}\phi+ 
        \epsilon_{\perp}\sin^{2}\phi
\label{eq:prob-trans}
\end{equation}

The expected coincidence rate, $R_{\phi}/R_{0}$, in the Freedman-Clauser
experiment~\cite{chsh:1969,horne:1970} predicted by QM is given by
Equation ~\ref{eq:QM-calc}.
\begin{eqnarray}
 R_{\phi}/R_{0} & = & \frac{1}{4}
     (\epsilon^{1}_{\parallel}+\epsilon^{1}_{\perp})
     (\epsilon^{2}_{\parallel}+\epsilon^{2}_{\perp})+  \nonumber \\ 
&   &     \frac{1}{4}(\epsilon^{1}_{\parallel}-\epsilon^{1}_{\perp})
          (\epsilon^{2}_{\parallel}-\epsilon^{2}_{\perp})
          F_{1}(\theta)\cos2\phi
\label{eq:QM-calc}
\end{eqnarray}
where the superscripts refer to the two different analyzers;
$F_{1}(\theta)$ is a function of the acceptance angle of the 
detectors; $R_{\phi}$ is the measured coincidence
rate with the analyzers at a relative angle $\phi$; and
$R_{0}$ is the measured coincidence rate with both of the
analyzers removed.  $R_{\phi}/R_{0}$ therefore measures
the effective coincidence rate and removes any effect
of the detector efficiencies. 
Putting in the transmission efficiencies of the analyzers 
and acceptance angle 
measured by Freedman and Clauser~\cite{freedman-clauser:1972},
the expected coincidence rate is given by

$$
R_{\phi}/R_{0}=0.2512+0.2124\times\cos2\phi
$$

The results of the Freedman-Clauser experiment confirmed this QM
calculation within the estimated errors of the experiment
for all nine relative analyzer angles considered. 

%MC simulation section
\section{MC simulation of the generic wave function collapse model
and a local realistic model}
\label{sec-copenhagen_and_LR_MC}
If the generic wave function collapse interpretation of QM is correct, 
a Monte Carlo (MC)
simulation of the Freedman-Clauser experiment should also
match this QM calculation.  To test this hypothesis
a MC simulation of the Freedman-Clauser experiment
was created using Mathematica.  Random polarization 
angles with a uniform distribution between $-\pi/2$
to $\pi/2$ were 
generated for the photons using the standard
random number generator within Mathematica.

The probability of photon 1 passing analyzer 1, $P(\phi)$ 
was calculated using Equation ~\ref{eq:prob-trans}
with a random polarization for $\phi$.
A random number, $N$, between 0 and the maximum
probability was then generated
and if $N < P(\phi)$, photon 1 was considered to
have been transmitted by analyzer 1 
(von Neumann's acceptance-rejection technique).  

Since photon 1 and photon 2 are entangled, it was
assumed, pursuant to the generic wave function collapse 
interpretation of QM,
that the measurement of the polarization of photon 1 collapsed
the wave function such that the polarization of
photon 2 was the same as the measured polarization
angle of photon 1 (i.e. the angle of analyzer 1). The same MC
procedure used for photon 1 was then used to determine
if photon 2 was transmitted by analyzer 2. If it was,
it was counted as a coincident measurement. 
To account for the effect of the acceptance angle of
the detector on the coincidence ratio ($R_{\phi}/R_{0}$)
a multiplicative coefficient, $F_2$ was applied to the 
coincidence ratio. $F_2$ was 
determined by fitting the generic wave function collapse model
MC simulation to the curve defined by the QM calculation. 

\begin{figure}
\includegraphics[width=8cm]{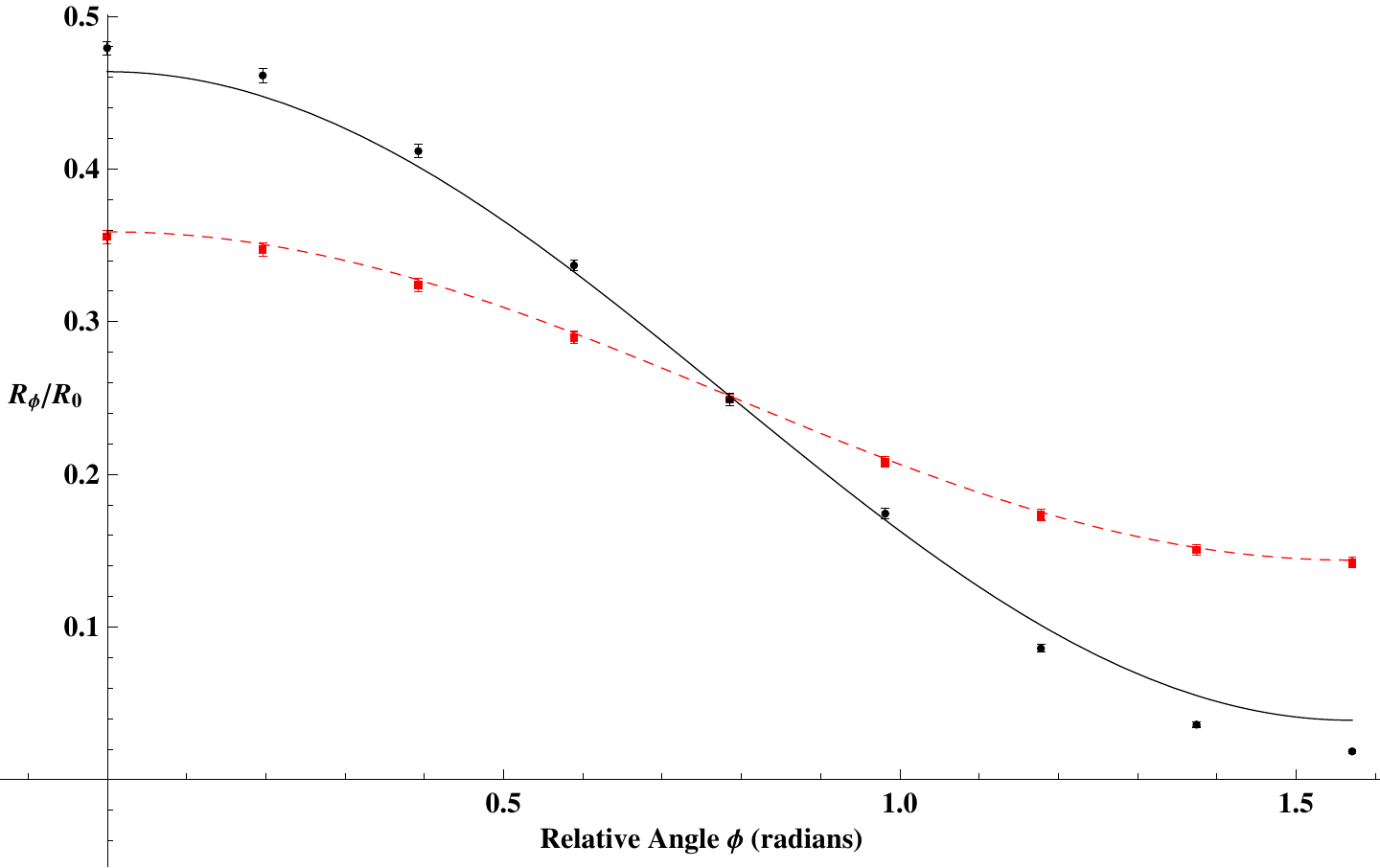}
\caption{The solid black line is the QM calculation, 
black data points are MC simulations
assuming the generic wave function collapse model; 
the dashed red line is the classical calculation assuming 
a local realistic model, 
the red data points (squares) are MC simulations
of a local realistic model. The error bars represent 
$\pm 3$ standard deviations.}
\label{fig:combo-plot}
\end{figure}

A MC run consisted of 100,000 simulated
photon pairs at each of the nine analyzer angles
used in the Freedman-Clauser experiment 
and 500 separate MC ``experiments'' were combined 
($5\times10^7$ total simulated photon pairs at each angle) 
to generate the solid black data points shown in 
Figure ~\ref{fig:combo-plot}.

% Combo Plot showing generic wave function collapse model
% and local realistic classical plot

The general shape of the curve closely matches the QM
calculation, but with disagreement of approximately $3\%$ 
(7 to 10 standard deviations) at 
analyzer angles less than or equal to $\pi/8$ and disagreement
between $15\%$ and $52\%$ (18 to 48 standard deviations) 
for analyzer angles greater than or equal to $3\pi/8$.  
The only free parameter
in the fit was the acceptance angle coefficient, $F_2$. The best
fit occurred with $F_2 = 0.9222 \pm 0.0002$.  
This value for $F_2$ was subsequently
used in all other MC simulations to account for the detector
acceptance. 

Since the QM calculation does not include an explicit assumption
regarding the polarizations of the two photons, it is possible
the physical correlation between them is either different than 
assumed here or simply undefined and the generic 
wave function collapse model 
should be considered merely an approximation of the underlying 
physical reality. To address the first possibility several other
physical correlations were tried ranging from no correlation at all
to fixed or random differences.  No alternative assumption
produced better agreement with the QM calculation. 

A local realistic model was specifically evaluated which
assumed the polarizations of the two photons were random but equal.
This physical interpretation is in good agreement with a classical
calculation using this assumption (Equation~\ref{eq:classic-calc} where
the variables have the same meanings as in Equation~\ref{eq:QM-calc}) 
but is in strong disagreement
with the results of the Freedman-Clauser experiment and the QM calculation.  
\begin{eqnarray}
 R_{\phi}/R_{0} & = & \frac{1}{4}
     (\epsilon^{1}_{\parallel}+\epsilon^{1}_{\perp})
     (\epsilon^{2}_{\parallel}+\epsilon^{2}_{\perp})+  \nonumber \\ 
&   &     \frac{1}{8}(\epsilon^{1}_{\parallel}-\epsilon^{1}_{\perp})
          (\epsilon^{2}_{\parallel}-\epsilon^{2}_{\perp})
          \cos2\phi
\label{eq:classic-calc}
\end{eqnarray}
The red, dashed curve on 
Figure ~\ref{fig:combo-plot}
shows the classical calculation (Equation~\ref{eq:classic-calc}
using the efficiency values obtained by Freedman and Clauser)
and the red, square data
points show the results of the MC simulation.  The same
MC procedures described previously were used in this simulation
including the same value of the detector acceptance coefficient, $F_2$,
determined in the generic wave function collapse model fit. Other fixed
correlations between the two photons were tried but none produced
results that better fit the QM calculation. 

\section{MC simulation of a Dynamic State Reduction/collapse 
Model (smeared polarization)}
\label{sec-fuzzy_MC}
We next tested an interpretation
of QM wherein physical quantities or ``elements of physical reality''
using EPR terminology, have no precise values and are fundamentally
indeterminate (i.e. smeared).  
We believe this situation approximates the physical interpretation
represented by dynamical state reduction/collapse 
models~\cite{ghirardi:1986,barchielli:1982,barchielli:1983,bedingham:2011}. 

To explore this possibility a MC simulation was developed
wherein neither photon 1 nor photon 2 had precise polarizations.
The polarization of photon 1 was again assumed to be random 
with a uniform distribution from $-\pi/2$ to $\pi/2$. 
The probability of photon 1 being transmitted by analyzer 1 was
calculated using the same technique used in the generic 
wave function collapse model simulation.  

If photon 1 was transmitted by analyzer 1, 
the polarization for photon 2 was assumed to
have a Gaussian distribution about the analyzer 1
angle. The standard deviation of this Gaussian distribution
is a function of the presumed inherent uncertainty or ``smearing''
of the photon's polarization.    
To determine if photon 2 was transmitted by analyzer 2, 
a random number was generated using this Gaussian distribution. 
This number was taken to be the effective polarization of photon 2.  
Whether or not photon 2 was transmitted by analyzer 2 was then
determined using the same procedure as in photon 1. 
If both photons were transmitted by their respective
analyzers a coincident detection was counted.  The same detector
acceptance coefficient, $F_2$, was applied to the coincidence rate.

\begin{figure}
\includegraphics[width=8cm]{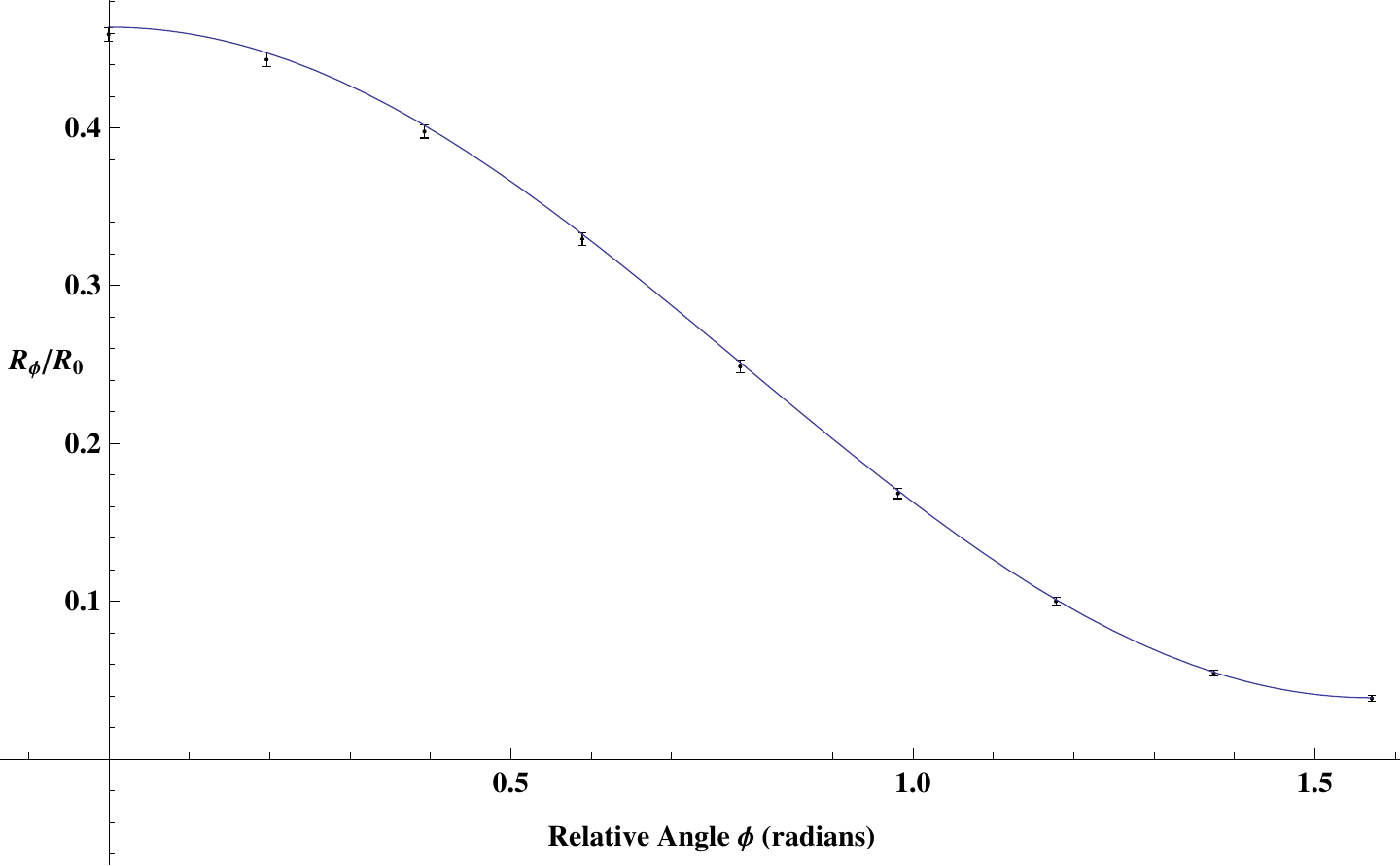}
\caption{The solid black line is the QM calculation, 
black data points are MC simulations
based on a smeared polarization model.  The error bars
represent $\pm 3$ standard deviations.}
\label{fig:fuzzymodel-plot}
\end{figure}

The Gaussian standard deviation that produced the best fit to the QM
calculation was $\sigma = 0.2131 \pm 0.0009$.  % +/- 3 sigma error 
Figure~\ref{fig:fuzzymodel-plot}
shows the results of the MC simulation (data points)
and the solid black line is the QM calculation. 
%
% Plot showing smeared polarization model
%
The smeared polarization model shown in 
Figure ~\ref{fig:fuzzymodel-plot} 
differs from the QM calculation by 
approximately $1\%$ (1 to 3 standard deviations)
for all nine analyzer angles. 
This difference could
be reduced to roughly $0.1\%$ by adjusting
the $F_2$ parameter by $1\%$.  
The mean chi squared for the generic wave function collapse
model was $0.021 \pm 0.002$ and $0.0003 \pm 0.0002$ %+/- 3 sigma errors
for the smeared polarization model 
which is almost a two orders of magnitude better fit
without adjusting the $F_2$ parameter. However, even with no
$F_2$ parameter adjustment the smeared polarization
model is a quasi-two variable fit so one would expect it
to better match the QM calculation.  The amount of
improvement, however, at least suggests that a
smeared polarization interpretation is a better model of
physical reality and appears to support dynamical state
reduction/collapse models.

The fit to the Gaussian standard deviation, $\sigma$, also appears
to support dynamical state reduction/collapse models. Dynamical state 
reduction/collapse models utilize a 
``localization distance''~\cite{ghirardi:1986,bassi:2003} 
or ``correlation length $(r_c)$'' ~\cite{adler:2007,bassietal:2012}, 
generally taken to be on the order of $r_c \simeq 10^{-5}$ cm.
Current experiments to test 
dynamical state reduction/collapse models
do not provide a bound on $r_c$~\cite{bassietal:2012}.  %bottom of page 90
 
The atomic cascade of calcium
used in the Freedman-Clauser experiment produced photons
with an average wavelength of $4.87\times 10^{-5}$ cm
($5513\AA$ and $4227\AA$ respectively for the two photons). 
Taking this as the ``size'' of the photons,
an uncertainty
in the polarization angle of $0.2131 \pm 0.0009$ radians equates to a
distance uncertainty of $(1.04 \pm 0.14) \times 10^{-5}$ cm in 
close agreement with the assumed value of $r_c$
used in dynamical state reduction/collapse models.



\end{document}